# Misalignment Resilient Diffractive Optical Networks

Deniz Mengu[1,2,3], Yifan Zhao[1,3], Nezih T. Yardimci[1,3], Yair Rivenson[1,2,3], Mona Jarrahi[1,3], Aydogan Ozcan[1,2,3,4,*]

[1] Electrical and Computer Engineering Department, University of California, Los Angeles, CA, 90095, USA
[2] Bioengineering Department, University of California, Los Angeles, CA, 90095, USA
[3] California NanoSystems Institute, University of California, Los Angeles, CA, 90095, USA
[4] Department of Surgery, David Geffen School of Medicine, University of California, Los Angeles, CA, 90095, USA.
* Corresponding author: ozcan@ucla.edu

**Abstract:** As an optical machine learning framework, Diffractive Deep Neural Networks (D²NN) take advantage of data-driven training methods used in deep learning to devise light-matter interaction in 3D for performing a desired statistical inference task. Multi-layer optical object recognition platforms designed with this diffractive framework have been shown to generalize to unseen image data achieving e.g., >98% blind inference accuracy for hand-written digit classification. The multi-layer structure of diffractive networks offers significant advantages in terms of their diffraction efficiency, inference capability and optical signal contrast. However, the use of multiple diffractive layers also brings practical challenges for the fabrication and alignment of these diffractive systems for accurate optical inference. Here, we introduce and experimentally demonstrate a new training scheme that significantly increases the robustness of diffractive networks against 3D misalignments and fabrication tolerances in the physical implementation of a trained diffractive network. By modeling the undesired layer-to-layer misalignments in 3D as continuous random variables in the optical forward model, diffractive networks are trained to maintain their inference accuracy over a large range of misalignments; we term this diffractive network design as *vaccinated* D²NN (v-D²NN). We further extend this vaccination strategy to the training of diffractive networks that use differential detectors at the output plane as well as to jointly-trained hybrid (optical-electronic) networks to reveal that all of these diffractive designs improve their resilience to misalignments by taking into account possible 3D fabrication variations and displacements during their training phase.

## 1 Introduction

Deep learning has been redefining the state-of-the-art for processing various signals collected and digitized by different sensors, monitoring physical processes for e.g., biomedical image analysis[1]–[4], speech recognition[5], [6] and holography[7]–[10], among many others[11]–[17]. Furthermore, deep learning and related optimization tools have been harnessed to find data-driven solutions for various inverse problems arising in, e.g., microscopy[18]–[22], nano-photonic designs and plasmonics[23]–[25]. These demonstrations and others have been motivating some of the recent advances in optical neural networks and related optical computing techniques that aim to exploit the computational speed, power-efficiency, scalability and parallelization capabilities of optics for machine intelligence applications[26]–[45].

Toward this broad goal, Diffractive Deep Neural Networks (D²NN)[36]–[39] have been introduced as a machine learning framework that unifies deep learning-based training of matter with the physical models governing light propagation to enable all-optical inference through a set of diffractive layers. The training stage of a diffractive network is performed using a computer, and relies on deep learning and error backpropagation methods to tailor the light-matter interaction across a set of diffractive layers that collectively perform a given machine learning task, e.g., object classification. Previous studies on D²NNs have demonstrated the generalization capability of these multi-layer diffractive network designs to new, unseen image data. For example, using a 5-layer diffractive network architecture, >98% and >90% all-optical blind testing accuracies have been reported [38] for the classification of the images of handwritten digits (MNIST) [46] and fashion products (Fashion-MNIST) [47] that are encoded in the amplitude and phase channels of the input plane, respectively. Successful experimental demonstrations of these all-optical classification systems have been reported using 3D-printed diffractive layers that conduct inference by modulating the incoming object wave at terahertz (THz) wavelengths.

Despite the lack of nonlinear optical elements in these previous implementations, diffractive optical networks have been shown to offer significant advantages in terms of (1) inference accuracy, (2) diffraction efficiency and (3) signal contrast, when the number of successive diffractive layers in the network design is increased [37]. A similar depth advantage was also demonstrated in [39], where instead of a statistical inference task such as image classification, the $D^2NN$ framework was utilized to solve an inverse design problem to achieve e.g., spatially-controlled wavelength de-multiplexing of a broadband source. While these multi-layer diffractive architectures offer significantly better performance for generalization and application-specific design merits, they also pose practical challenges for the fabrication and opto-mechanical assembly of these trained diffractive models.

Here, we present a training scheme that substantially increases the robustness of diffractive optical networks against physical misalignments and fabrication tolerances. Our scheme models and introduces these undesired system variations and layer-to-layer misalignments as continuous random variables during the deep learning-based training of the diffractive model to significantly improve the error tolerance margins of diffractive optical networks; this process of introducing random misalignments during the training phase will be termed as *vaccination* of the diffractive network, and the resulting designs will be referred to as vaccinated $D^2NNs$ (v-$D^2NNs$). To demonstrate the efficacy of our strategy, we trained diffractive network models composed of 5 diffractive layers for all-optical classification of hand-written digits, where we utilized in the training phase independent and uniformly distributed displacement/misalignment vectors for x, y, and z directions of each diffractive layer. Our results indicate that v-$D^2NN$ framework enables the design of diffractive optical networks that can maintain their object recognition performance against severe layer-to-layer misalignments, providing nearly flat blind inference accuracies within the displacement/misalignment range adopted in the training.

To experimentally demonstrate the success of v-$D^2NN$ framework we also compared two 3D-printed diffractive networks, each with 5 diffractive layers that were designed for hand-written digit classification under monochromatic THz illumination ($\lambda = \sim0.75$ mm): the first network model was designed without the presence of any misalignments (non-vaccinated) and the second one was designed as a v-$D^2NN$. After the fabrication of each diffractive network, the $3^{rd}$ diffractive layer was on purpose misaligned to different 3D positions around its ideal location. The experimental results confirmed our numerical analysis to reveal that the v-$D^2NN$ design can preserve its inference accuracy despite a wide range of physical misalignments, while the standard $D^2NN$ design frequently failed to recognize the correct data class due to these purposely-introduced misalignments.

We also combined our v-$D^2NN$ framework with the differential diffractive optical networks [38] and the jointly-trained optical-electronic (hybrid) neural network systems. Differential diffractive classification systems assign a pair of detectors (generating one positive and one negative signal) for each data class to mitigate the strict non-negativity constraint of optical intensity, and were demonstrated to offer superior inference accuracy compared to standard diffractive designs [38]. When trained against misalignments using the presented v-$D^2NN$ framework, differential diffractive networks are also shown to preserve their performance advantages for all-optical classification. However, both differential and standard diffractive networks fall short in matching the adaptation capabilities of a hybrid diffractive network system that uses a modest, single-layer fully-connected architecture with only 110 learnable parameters in the electronic domain, following the diffractive optical front-end.

In addition to misalignment related errors, the presented vaccination framework can also be adopted to mitigate other error sources in diffractive network models, e.g., detection noise and fabrication imperfections or artefacts, provided that the approximate analytical models and the probability distributions of these factors are utilized during the training stage. We anticipate that v-$D^2NNs$ will be the gateway of diffractive optical networks and the related hybrid neural network schemes towards practical machine vision and sensing applications, by mitigating various sources of error between the training forward models and the corresponding physical hardware implementations.

# 2   Results

Figure 1 illustrates three different types of diffractive optical network-based object recognition systems investigated in this work. We focused on 5-layer diffractive optical network architectures as shown in Fig. 1 that are fully-connected, meaning that the half cone angle of the secondary wave created by the diffractive features (neurons) of size, e.g., $\delta=0.53\lambda$, is large enough to enable communication between all the features on two successive diffractive layers that are placed e.g., 40λ apart in axial direction. On the transverse plane, each diffractive layer extends from -100×δ to 100×δ on x and y directions around the optical axis, and therefore the edge length of each diffractive surface in total is 200×δ (~106.66λ). With this outlined diffractive network architecture, the standard D²NN training routine updates the trainable parameters of the diffractive layers at every iteration based on the mean gradient computed over a batch of training samples with respect to a loss function, specifically tailored for the desired optical machine learning application, e.g., cross-entropy for supervised object recognition systems [37], until a convergence criterion is satisfied. Since this conventional training approach assumes perfect alignment throughout the training, the sources of statistical variations in the resulting model are limited to the initial condition of the diffractive network parameter space and the sequence of the training data introduced to the network.

## 2.1 Training and testing of v-D²NNs

The training of vaccinated diffractive optical networks mainly follows the same steps as the standard D²NN framework; except, it additionally incorporates system errors, e.g. misalignments, based on their probability distribution functions into the optical forward model. In this work, we modelled each orthogonal component of the undesired 3D displacement vector of each diffractive layer, $D = (D_X, D_Y, D_Z)$, as uniformly distributed, independent random variables as follows;

$$D_X \sim U(-\Delta_X, \Delta_X) \tag{1a},$$

$$D_y \sim U(-\Delta_y, \Delta_y) \tag{1b},$$

$$D_Z \sim U(-\Delta_Z, \Delta_Z) \tag{1c},$$

where $\Delta*$ denotes the shift along the corresponding axis, (*), reflecting the uncertainty in our physical assembly/fabrication of the diffractive model. During the training, the random displacement vector of *each* diffractive layer, $\mathbf{D}$, takes different values sampled from the probability distribution of its components, $D_X$, $D_Y$ and $D_Z$, for each batch of training samples. Consequently, the location of layer $l$ at $i^{th}$ iteration/batch, $\mathbf{L^{(l,i)}}$, can be expressed as;

$$\mathbf{L^{(l,i)}} = (L_x^l, L_y^l, L_z^l) + (D_X^{(l,i)}, D_y^{(l,i)}, D_Z^{(l,i)}) \tag{2}$$

where the first and the second vectors on the right-hand side denote the ideal location of the diffractive layer $l$, and a random realization of the displacement vector, $\mathbf{D^{(l,i)}}$, of layer $l$ at the training iteration $i$, respectively. The displacement vector of each layer is independently determined, i.e., each layer of a diffractive network model can move within the displacement ranges depicted in Eq. (1) without any dependence on the locations of the other diffractive layers.

Opto-mechanical assembly and fabrication systems, in general, use different mechanisms to control the lateral and axial positioning of optical components. Therefore, we split our numerical investigation of the vaccination process into two: the lateral and axial misalignment cases. For the vaccination of diffractive optical network models against layer-to-layer misalignments on the transverse plane, we assumed $D_X$ and $D_y$ are i.i.d random variables during the training, i.e. they are independent with a parameter of $\Delta_X = \Delta_y = \Delta_{tr}$, and $D_Z$ was set to be 0. The axial case, on the other hand, sets $\Delta_{tr}$ to be 0 throughout the training leaving $D_Z \sim U(-\Delta_{(z,tr)}, \Delta_{(z,tr)})$ as the only source of inter-layer misalignments.

Following a similar path with the training, the blind testing of the presented diffractive network models updates the random displacement vector of each layer $l$, $\mathbf{D^{(l,m)}}$, for *each* test sample $m$. The reported accuracies throughout our analyses reflects the *blind testing accuracies* computed over the 10K image test set of MNIST digits where *each test sample propagates through a diffractive network model that experiences a different realization of the random variables* depicted in Eq. (1) for each diffractive layer, i.e. *there are 10K different configurations that a diffractive network model was misaligned throughout the testing stage*. Furthermore, similar to the training process, during the blind testing against lateral misalignments, it was assumed that $D_X$ and $D_y$ are i.i.d random variables with $\Delta_X = \Delta_y = \Delta_{test}$, and similarly, the axial displacements or misalignments were determined by $D_Z \sim U(-\Delta_{(z,test)}, \Delta_{(z,test)})$.

## 2.2   Misalignment analysis of all-optical and hybrid diffractive systems

Figures 2A and 2D illustrate the blind testing accuracies provided by the standard diffractive optical network architecture (Fig. 1A) trained against various levels of undesired axial and lateral misalignments, respectively. Focusing on the testing accuracy curve obtained by the error-free design (dark blue) in Figs. 2A and 2D, it can be noticed that the diffractive optical networks are more susceptible to lateral misalignments compared to axial misalignments. For instance, when $\Delta_{test}$ is taken as 2.12λ, inducing random lateral fluctuations on each diffractive layer's location around the optical axis, the blind testing accuracy achieved by the non-vaccinated standard diffractive optical network decreases to 38.40% from 97.77% (obtained in the absence of misalignments). As we further increase the level of lateral misalignments, the error-free diffractive optical network almost completely loses its inference capability by achieving, e.g. 19.24% blind inference accuracy for $\Delta_{test}$=4.24λ (i.e., the misalignment range in each lateral direction of a diffractive layer is -8δ to 8δ). On the other hand, when the diffractive layers are randomly misaligned on the longitudinal direction alone, the inference performance does not drop as excessively as the lateral misalignment case; for example, even when $\Delta_{(z,test)}$ becomes as large as 19.2λ, the error-free diffractive network manages to obtain an inference accuracy of 49.8%.

As demonstrated in Fig. 2D, the rapid drop in the testing accuracy of diffractive optical classification systems under physical misalignments can be mitigated by using the v-D²NN framework. Since v-D²NN training introduces displacement errors in the training stage, the diffractive optical networks can adopt to those variations preserving their inference performance over large misalignment margins. As an example, the 38.40% blind testing accuracy achieved by the non-vaccinated diffractive design with a lateral misalignment range of $\Delta_{test}$=2.12λ, can be increased to 94.44% when the same architecture is trained with a similar error range using the presented vaccination framework (see the purple line in Fig. 2D). On top of that, the vaccinated design does not compromise the performance of the all-optical object recognition systems when the ideal conditions are satisfied. Compared to the 97.77% accuracy provided by the error-free design, this new vaccinated network (purple line in Fig. 2D) obtains 96.1% in the absence of misalignments. In other words, the ~56% inference performance gain of the vaccinated diffractive network under physical misalignments comes at the expense of only 1.67% accuracy loss when the opto-mechanical assembly perfectly matches the numerical training model. In case the level of misalignment-related imperfections in the fabrication of the diffractive network is expected to be even smaller, one can design improved v-D²NN models that achieve e.g., 97.38%, which corresponds to only 0.39% inference accuracy loss compared to the error-free models at their peak (perfect alignment case) while at the same time providing >4% blind testing accuracy improvement under mild misalignment, i.e., $\Delta_{test}$ =0.53λ. Similarly, when we compare the blind inference curves of the error-free and vaccinated network designs in Fig. 2A, one can notice that the v-D²NN framework can easily recover the performance of the diffractive digit classification networks in the case where the displacement errors are restricted to be on the longitudinal axis. For example, with $\Delta_{(z,test)}$=2.4λ, the inference accuracy of the error-free diffractive network (dark blue) is reduced to 94.88%, while a vaccinated diffractive network that was already trained against the same level of misalignment, $\Delta_{(z,tr)}$=2.4λ (yellow), retains 97.39% blind inference accuracy under the same level of axial misalignment.

Next, we combined our v-D²NN framework with the *differential* diffractive network architecture: the blind testing results of various differential handwritten digit recognition systems under axial and lateral misalignments are reported in Figs. 2B and Fig. 2E, respectively. Figure 3 also provides a direct comparison of the blind inference accuracies of these two all-optical diffractive machine learning architectures under different levels of misalignments. Figs. 3A and 3G compare the error-free designs of differential and standard diffractive network architectures, which reveal that although the differential design achieves slightly better blind inference accuracy, 97.93%, in the absence of alignment errors, as soon as the misalignments reach beyond a certain level, the performance of a differential design decreases faster than the standard diffractive network. This means that they are more vulnerable against the system variations that they were not trained against. Since the number of detectors inside an output region-of-interest is twice as many in differential diffractive networks compared to the standard diffractive network architecture (see Fig. 1A-B), the detector signals are more prone to have cross-talk when the diffractive layers are experiencing uncontrolled mechanical displacements. With the introduction of vaccination during the training phase, however, differential diffractive network models can adapt to these system variations as in the case of standard diffractive optical networks. Compared to standard diffractive optical networks, the differential counterparts that are vaccinated generate higher inference accuracies when the misalignment levels are small. In Fig. 3H, for instance, the vaccinated differential design (red curve) achieves 97.3% blind inference accuracy while the vaccinated standard diffractive network (blue curve) can provide 96.91% for the case $\Delta_{test}$ = $\Delta_{tr}$ = 0.53λ. In Fig. 3I, where the vaccination range on x and y axis is twice as large compared to Fig. 3H, the differential network reveals the correct digit classes with an accuracy of 96.18% when it is tested at an equal displacement/misalignment uncertainty to its vaccination level; on the other hand, the standard diffractive network can achieve 95.79% under the same training and testing conditions. Beyond this level of misalignment, the differential systems slowly lose their performance advantage and the standard diffractive networks starts to perform on par with their differential counterparts. One exception to this behavior is shown in Fig. 3K, where the misalignment range of the diffractive layers during the training causes cross-talk among the differential detectors at a level that hurts the evolution of the differential diffractive network, leading to a consistently worse inference performance compared to the standard diffractive design. A similar effect also exists for the case illustrated in Fig. 3L; however, this time, the standard diffractive optical network design also experiences a similar level of cross-talk among the class detectors at the output plane. Therefore, as demonstrated in Fig. 3L, the differential diffractive optical network recovers its performance gain thriving over the standard diffractive network design with a higher optical classification accuracy. This performance gain of the differential design depicted in Fig. 3L, can be translated to the smaller misalignment cases, e.g., $\Delta_{test} = \Delta_{tr} = 4.24\lambda$, simply by increasing the distance between the detectors at the output plane for differential diffractive optical network designs, i.e. setting the region-of-interest covering the detectors to be larger compared to the standard diffractive network architecture.

Figure 3 also outlines a comparison of the differential and standard diffractive all-optical object recognition systems against hybrid diffractive neural networks under various levels of misalignments. For the hybrid neural network models presented here, we jointly trained a 5-layer diffractive optical front-end and a single-layer fully-connected electronic network, communicating through discrete detectors at the output plane. To provide a fair comparison with the all-optical diffractive systems, we used 10 discrete detectors at the output plane of these hybrid configurations, same as in the standard diffractive optical network designs (see Figs. 1A and 1C). The blind inference accuracies obtained by these hybrid neural network systems under different levels of misalignments are shown in Figs. 2C and 2F. When the opto-mechanical assembly of the diffractive network is perfect, the error-free, jointly-optimized hybrid neural network architecture can achieve 98.3% classification accuracy surpassing the all-optical counterparts as well as the all-electronic performance of a single-layer fully-connected network, which achieves 92.48% classification accuracy using >75-fold more trainable connections without the diffractive optical network front-end. As the level of misalignments increases, however, the error-free hybrid network fails to maintain its performance and its inference accuracy quickly falls. The v-D²NN framework helps the hybrid neural systems during the joint evolution of the diffractive and the electronic networks and makes them resilient to misalignments. For example, the handwritten digit classification accuracy values presented for the standard diffractive networks in Fig. 3H (96.91%) and Fig. 3I (95.79%) have improved to 97.92% and 97.15%, respectively, for the hybrid neural network system (yellow curve), indicating ~1% accuracy gain over the all-optical models under the same level of misalignment (i.e., 0.53λ for Fig. 3H and 1.06λ for Fig. 3I). As the level of misalignments in the diffractive optical front-end increases, the cross-talk between the detectors at the output plane also increases. However, for a hybrid network design there is no *direct* correspondence between the data classes and the output detectors, and therefore the joint-training under the vaccination scheme introduced in this work directs the evolution of the electronic network model accordingly and opens up the performance gap further between the all-optical diffractive classification networks and the hybrid systems as illustrated in Figs. 3K and 3L. A similar comparative analysis, along the lines of Figs. 2 and 3, is also conducted for phase-encoded input objects (Fashion-MNIST dataset), which is reported in Supplementary Figs. S4 and S5.

## 2.3 Experimental results

The error-free standard diffractive network design that achieves 97.77% blind inference accuracy for the MNIST dataset as presented in Figs. 2A, 2D, 3A and 3G, offers a power efficiency of ~0.07% on average over the blind testing samples (see Supplementary Information for details). This relatively low power efficiency is mostly due to the absorption of our 3D printing material at THz band. Specifically, ~88.62% of the optical power right after the object is absorbed by the 5 diffractive layers, while 11.17% is scattered around during the light propagation. Due to the limited optical power in our THz source and the noise floor of our detector, we trained an error-free standard diffractive optical network model with a slightly compromised digit classification performance for the experimental verification of our v-D²NN framework. This new error-free diffractive network provides a blind inference accuracy of 97.19%, and it obtains ~3× higher power efficiency of ~0.2%. In addition to improved power efficiency, this new diffractive network model with 97.19% classification accuracy also achieves ~10× better signal contrast (ψ) [37] between the optical signal collected by the detector corresponding to the true object label and its closest competitor, i.e. the second maximum signal (see Supplementary Information for details). The layers of this error-free diffractive network are shown in Fig. 1E. In addition, the comparison between the error-free, high-contrast standard diffractive optical network model and its lower contrast, lower efficiency counterpart in terms of their inference performance under misalignments is reported in Supplementary Fig. S1A.

Following the same power-efficient design strategy, we trained another diffractive optical network that is *vaccinated* against *both the lateral and axial misalignments* with the training parameters ($\Delta_{tr}$, $\Delta_{(z,tr)}$) taken as (4.24λ, 4.8λ). As in the case of the error-free design, the inference accuracy of this new vaccinated diffractive network shown in Fig. 5A is also compromised compared to the standard diffractive networks presented in Fig. 2D and Fig. 3K since it was trained to improve power efficiency and signal contrast. This design can achieve 89.5% blind classification accuracy for handwritten digits under ideal conditions, with the diffractive layers reported in Fig. 1D. A comprehensive comparison of the blind inference accuracies of the vaccinated diffractive networks shown in Figs. 2 and 3 and their high-contrast, high-efficiency counterparts are reported in Supplementary Fig. S1B.

The experimental verification of our v-D²NN framework was based on the comparison of the vaccinated and the error-free standard diffractive optical network designs in terms of the accuracy of their optical classification decisions under inter-layer misalignments. To this end, we fabricated the diffractive layers of the non-vaccinated and the vaccinated networks shown in Figs. 1D-E using 3D printing. The fabricated diffractive networks are depicted in Figs. 5C-D. In addition, we fabricated 6 MNIST digits selected from the blind testing dataset that are numerically correctly classified by both the vaccinated and the non-vaccinated diffractive network models without any misalignments. For a fair comparison, we grouped the correctly classified handwritten digits based on the signal contrast statistics provided by the non-vaccinated design. With $\mu_{SC}$, $\sigma_{SC}$ denoting the mean and the standard deviation of the signal contrast generated by the error-free diffractive network over the correctly classified blind testing MNIST digits, we selected 2 handwritten digits (Set 1) that satisfies the condition $\mu_{SC} + \sigma_{SC} < \{\psi, \psi'\} < \mu_{SC} + 2\sigma_{SC}$, where $\psi$ and $\psi'$ denote the signal contrasts created by the error-free and the vaccinated designs for a given input object, respectively. The condition on $\psi$ and $\psi'$ for the second set of 3D printed handwritten digits (Set 2), on the other hand, is slightly less restrictive, $\mu_{SC} < \{\psi, \psi'\} < \mu_{SC} + \sigma_{SC}$. By using this outlined approach, we selected 6 experimental test objects in total that are equally favorable for both the vaccinated and non-vaccinated diffractive networks.

To test the performance of the error-free and vaccinated diffractive network designs under different levels of misalignments, we shifted the 3ʳᵈ layer of both diffractive systems to 12 different locations around its ideal location as depicted in Fig. 5B. The perturbed locations of the 3ʳᵈ diffractive layer covers 4 different spots on each orthogonal direction. The distances between these locations are 1.2mm (1.6λ) along x and y, and 2.4mm (3.2λ) along z axes. These shifts cover a total length of 6.4λ (12 times the smallest feature size) along (x,y) and 12.8λ (0.32×40λ) along z axis, respectively.

Figure 5E shows a schematic of our THz setup that was used to test these diffractive networks and their misalignment performances (see Supplementary Information). Figure 6 reports the experimentally obtained optical signals for a handwritten digit '0' from Set 1 and a handwritten digit '5' from Set 2, received by the class detectors at the output plane based on the 13 different locations of the 3ʳᵈ diffractive layer of the vaccinated and the error-free networks. The first thing to note is that both the vaccinated and non-vaccinated networks can classify the two digits correctly when the 3ʳᵈ layer is placed at its ideal location within the set-up. As illustrated in Fig. 6A, as we perturb the location of the 3ʳᵈ layer, the error-free diffractive network fails at 9 locations while the vaccinated network correctly infers the object label at all the 13 locations for the handwritten digit '0'. In addition, the vaccinated network maintains its perfect record of experimental inference for the digit '5' despite the inter-layer misalignments as depicted in Fig. 6B. The error-free design, on the other hand, fails at 2 different locations of its 3ʳᵈ layer misalignment (see Fig. 6B). The experimental results for the remaining 4 digits are presented in Supplementary Figs. S2 and S3, confirming the same conclusions. In our experiments, all the objects were correctly classified when the 3ʳᵈ layer was placed at its ideal location. Out of the remaining 72 measurements (6 objects × 12 shifted/misaligned locations of the 3ʳᵈ layer), the error-free design failed to infer the correct object class in 23 cases, while the vaccinated network failed only 2 times, demonstrating its robustness against a wide range of misalignments as intended by the v-D²NN framework.

# 3 Discussion

As an example of a severe case of lateral misalignments, we investigated a scenario where each diffractive layer can move within the range (-8.48λ,8.48λ) around the optical axis in x and y directions. As demonstrated in Fig. 2D and Fig. 3G, when the error-free design (dark blue) is exposed to such large lateral misalignments, it can only achieve 12.8% test accuracy, i.e., it barely surpasses random guessing of the object classes. A diffractive optical network that is vaccinated against the same level of uncontrolled layer movement can partially recover the inference performance providing 67.53% blind inference accuracy. As the best performer, the hybrid neural network system composed of a 5-layer diffractive optical network and a single-layer fully-connected network can take this accuracy value up to 79.6% under the same level of misalignments, within the range (-8.48λ,8.48λ) for both x and y direction of each layer. When we compare the total allowed displacement range of each layer within the diffractive network (i.e., 16.96λ in each direction) and the size of our diffractive layers (106.66λ), we can see that they are quite comparable. If we imagine a lens-based optical imaging system and an associated machine vision architecture, in the presence of such serious opto-mechanical misalignments, this system would also fail due to acute aberrations substantially decreasing the image quality and the resolution. Our main motivation to include this severe misalignment case in our analyses was to test the limits of the adaptability of our vaccinated systems.

Figures 4A-B further summarize the inference accuracies of the differential diffractive networks and hybrid neural network systems at discrete points sampled from the corresponding curves depicted in Figs. 3G-L. In Fig. 4A, the best inference accuracy is achieved by the error-free (non-vaccinated) differential diffractive network model under perfect alignment of its layers. However, its performance drops in the presence of an imperfect opto-mechanical assembly. The vaccinated, diffractive all-optical classification networks provide major advantages to cope with the undesired system variations achieving higher inference accuracies despite misalignments. The joint-training of hybrid systems that are composed of a diffractive optical front-end and a single-layer electronic network (back-end) can adapt to un-controlled mechanical perturbations achieving higher inference accuracies compared to all-optical image classification systems. These results further highlight that, operating with only a few discrete opto-electronic detectors at the output plane, the D²NN-based hybrid architectures offer unique opportunities for the design of low-latency, power-efficient and memory-friendly machine vision systems for various applications.

In conclusion, we presented a design framework that introduces the use of probabilistic layer-to-layer misalignments during the training of diffractive neural networks to increase their robustness against physical misalignments. Beyond misalignments or displacements of diffractive layers, the presented vaccination framework can also be harnessed to decrease the sensitivity of diffractive optical networks to various error sources, e.g. detection noise or fabrication defects. We believe the presented training strategy will find use in the design of diffractive optical network-based machine vision and sensing systems, spanning different applications.

# Acknowledgements

The Ozcan Research Group at UCLA acknowledges the support of Fujikura, Japan.

# Figures

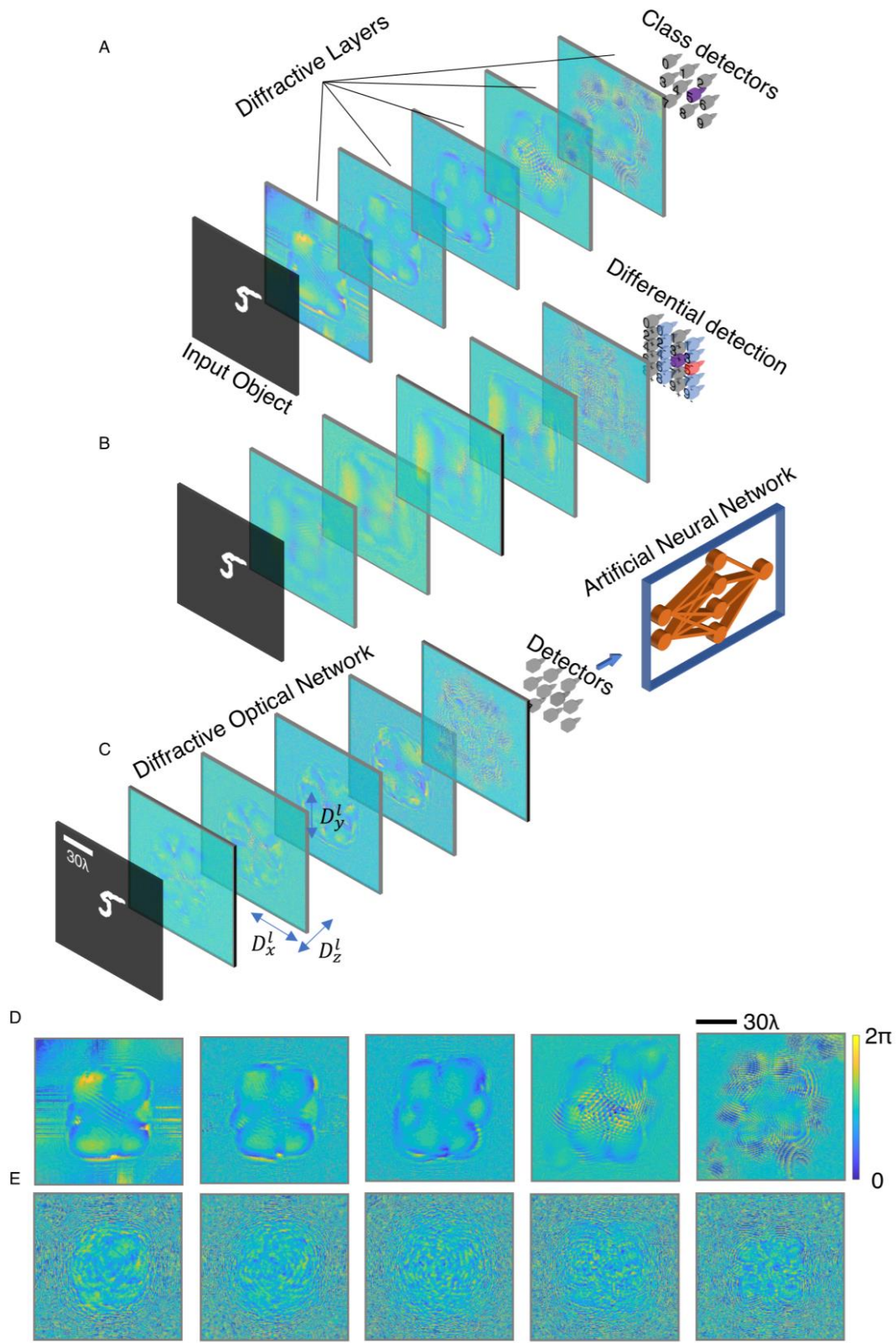

**Fig. 1: Different types of D²NN-based image classification systems. A** Standard D²NN framework trained for all-optical classification of handwritten digits. Each detector at the output plane represents a data class. **B** Differential D²NN trained for all-optical classification of handwritten digits. Each data class is represented by a pair of detectors at the output plane, where the normalized difference between these detector pairs represents the class scores. **C** Jointly-trained hybrid (optical-electronic) network system trained for classification of handwritten digits. The optical signals collected at the output detectors are used as inputs to the electronic neural network at the back-end, which is used to output the final class scores.

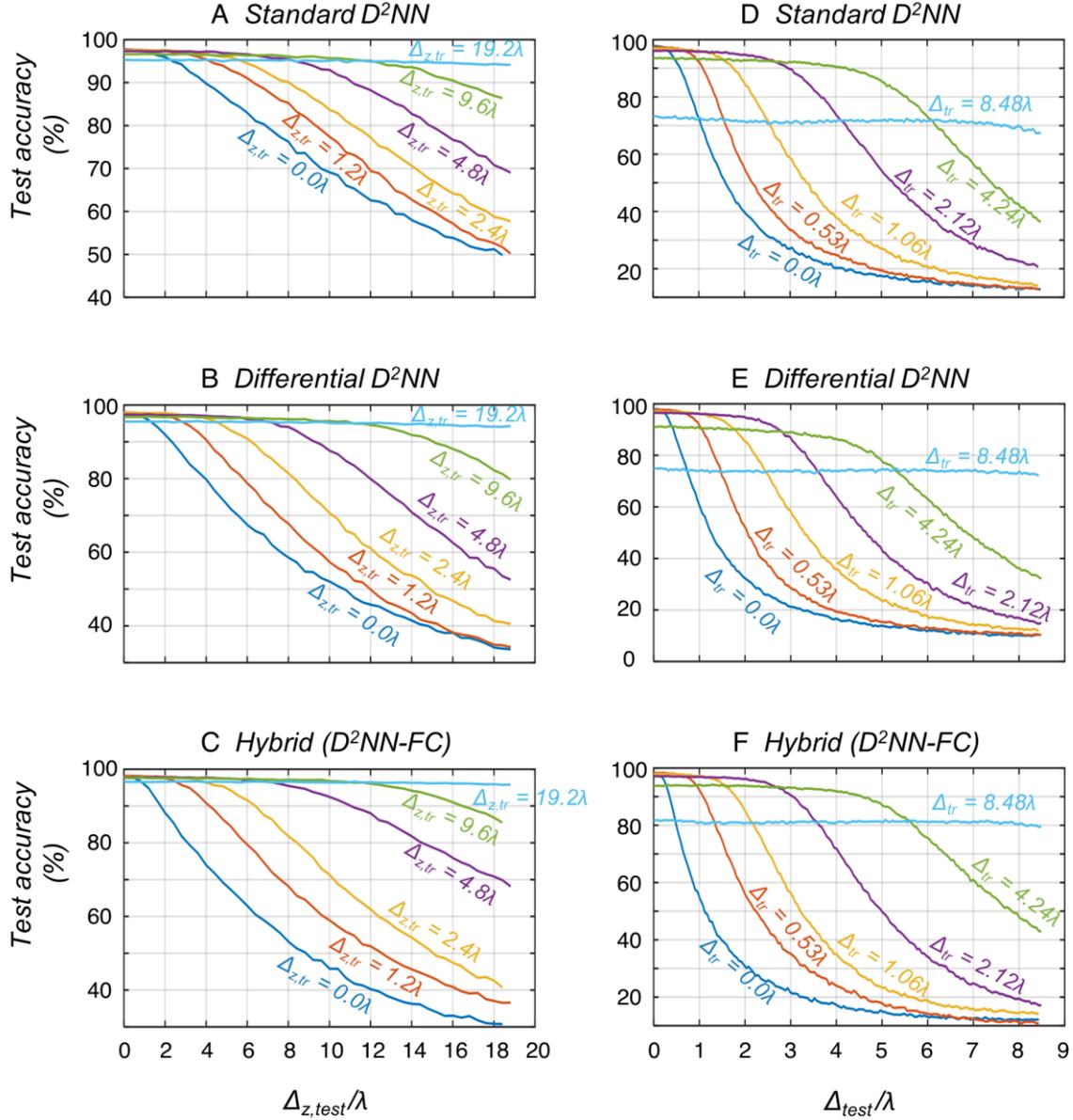

**Fig. 2: The sensitivity of the blind inference accuracies of different types of D²NN-based object classification systems against various levels of misalignments. A** Standard D²NN systems trained for all-optical handwritten digit classification with and without vaccination were tested against various levels of *axial* misalignments, determined by $\Delta_{z,test}$. **B** Same as A, except for differential D²NN architectures. **C** Same as A and B, except for hybrid (D²NN-FC) systems comprised of a jointly-trained 5-layer D²NN optical front-end and a single-layer fully-connected neural network at the electronic back-end, combined through 10 discrete opto-electronic detectors (see Fig. 1C). The comparison of these blind testing results reveals that as the axial misalignment increases during the training, $\Delta_{z,tr}$, the inference accuracy of these machine vision systems decrease slightly but at the same time they are able to maintain their performance over a wider range of misalignments during the blind testing, $\Delta_{z,test}$. **D** Standard D²NN systems trained for all-optical handwritten digit recognition with and without vaccination were tested against various levels of *lateral* misalignment levels, determined by $\Delta_{test}$. **E** Same as D except for differential D²NNs architectures. **F** Same as E and F, except for hybrid object recognition systems comprised of a jointly-trained 5-layer D²NN optical front-end and a single-layer fully-connected neural network at the electronic back-end, combined through 10 discrete opto-electronic detectors. The proposed vaccination-based training strategy improves the resilience of these diffractive networks to uncontrolled *lateral* and *axial* displacements of the diffractive layers with a modest compromise of the inference performance depending on the misalignment range used in the training phase.

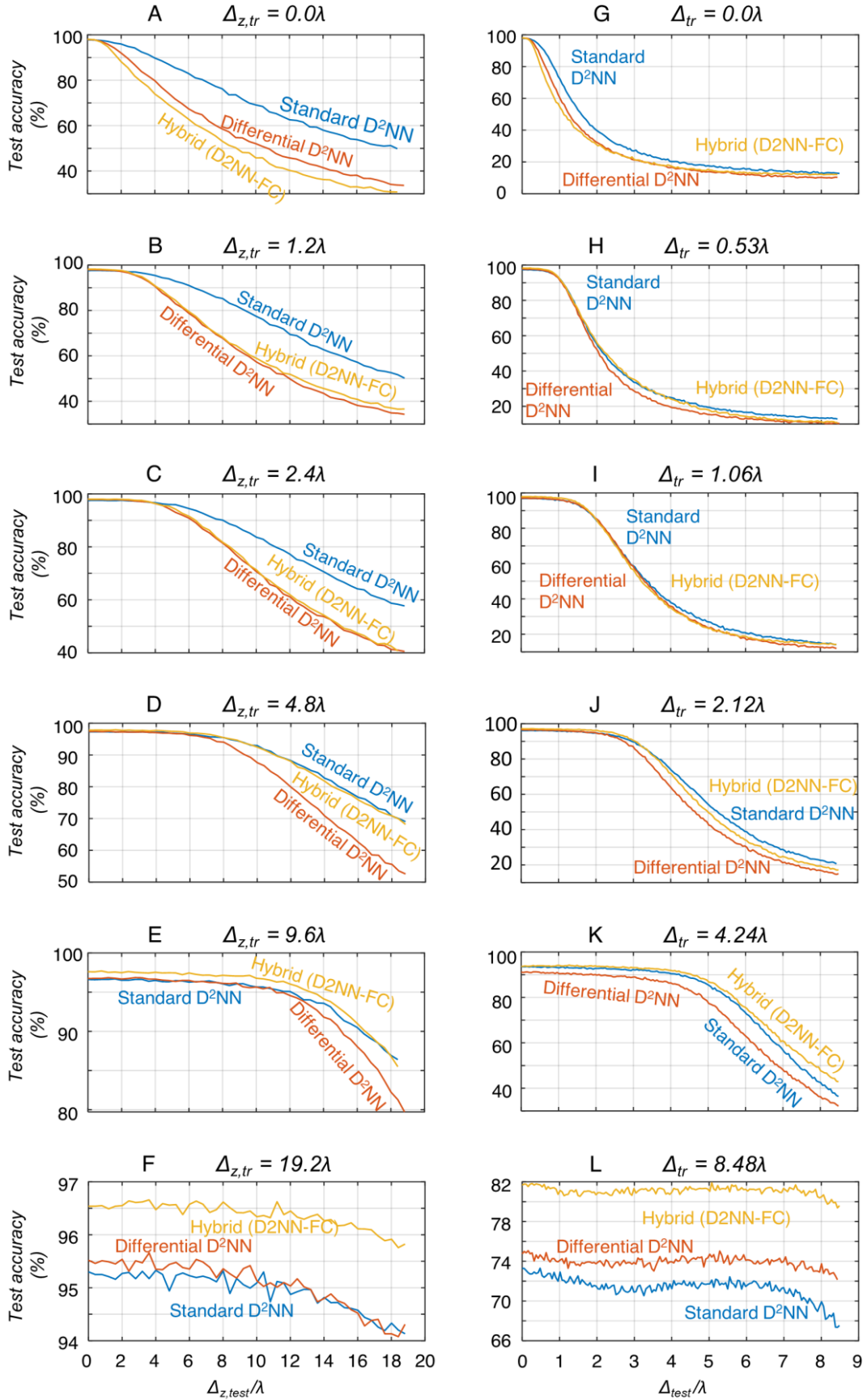

**Fig. 3: Comparison of different types of D²NN-based object classification systems trained with the same range of misa-lignments. A** Comparison of error-free designs, $\Delta_{z,tr} = 0.0\lambda$, for standard (blue), differential (red) and hybrid (yellow) object classification systems against different levels of *axial* misalignments, $\Delta_{z,test}$. **B** Comparison of standard (blue), differential (red)

and hybrid (yellow) object classification systems against different levels of *axial* misalignments when they are trained with $\Delta_{Z,tr}$ = 1.2λ. **C,D,E and F** are same as B, except during the training of the diffractive models the axial misalignment ranges are determined by $\Delta_{Z,tr}$, taken as 2.4λ, 4.8λ, 9.6λ and 19.2λ, respectively. **G** Comparison of error-free designs, $\Delta_{tr}$ = 0.0λ, for standard (blue), differential (red) and hybrid (yellow) object recognition systems against different levels of *lateral* misalignments, $\Delta_{test}$. **H** Comparison of standard (blue), differential (red) and hybrid (yellow) object classification systems against different levels of *lateral* misalignments when they are trained with $\Delta_{tr}$ = 0.53λ. **I,J,K and L** are same as H, except the *lateral* misalignment ranges during the training are determined by $\Delta_{tr}$, taken as 1.06λ, 2.12λ, 4.24λ and 8.48λ, respectively.

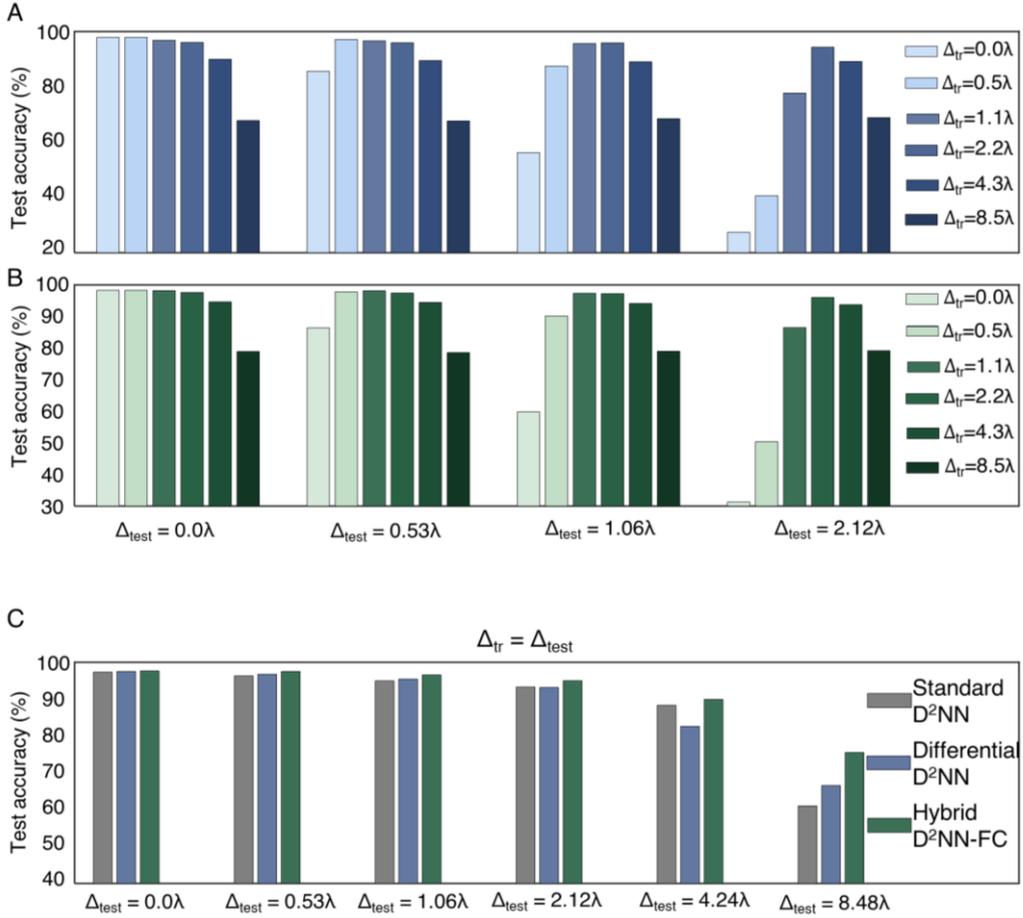

**Fig. 4: Summary of the numerical results for vaccinated D²NNs. A** The inference accuracy of the non-vaccinated ($\Delta_{tr}$ = 0.0λ) and the vaccinated ($\Delta_{tr} > 0.0λ$) differential D²NN systems trained for all-optical handwritten digit recognition quantified at different levels of testing misalignment ranges. The v-D²NN framework allows the all-optical classification systems to preserve their inference performance over a large range of misalignments. **B** Same as A, except for hybrid (D²NN-FC) systems comprised of a jointly-trained 5-layer D²NN optical front-end and a single-layer fully-connected neural network at the electronic back-end combined through 10 discrete opto-electronic detectors (see Fig. 1C). **C** Vaccination comparison of 3 diffractive network-based machine learning architectures depicted in Fig. 1; $\Delta_{tr} = \Delta_{test}$.

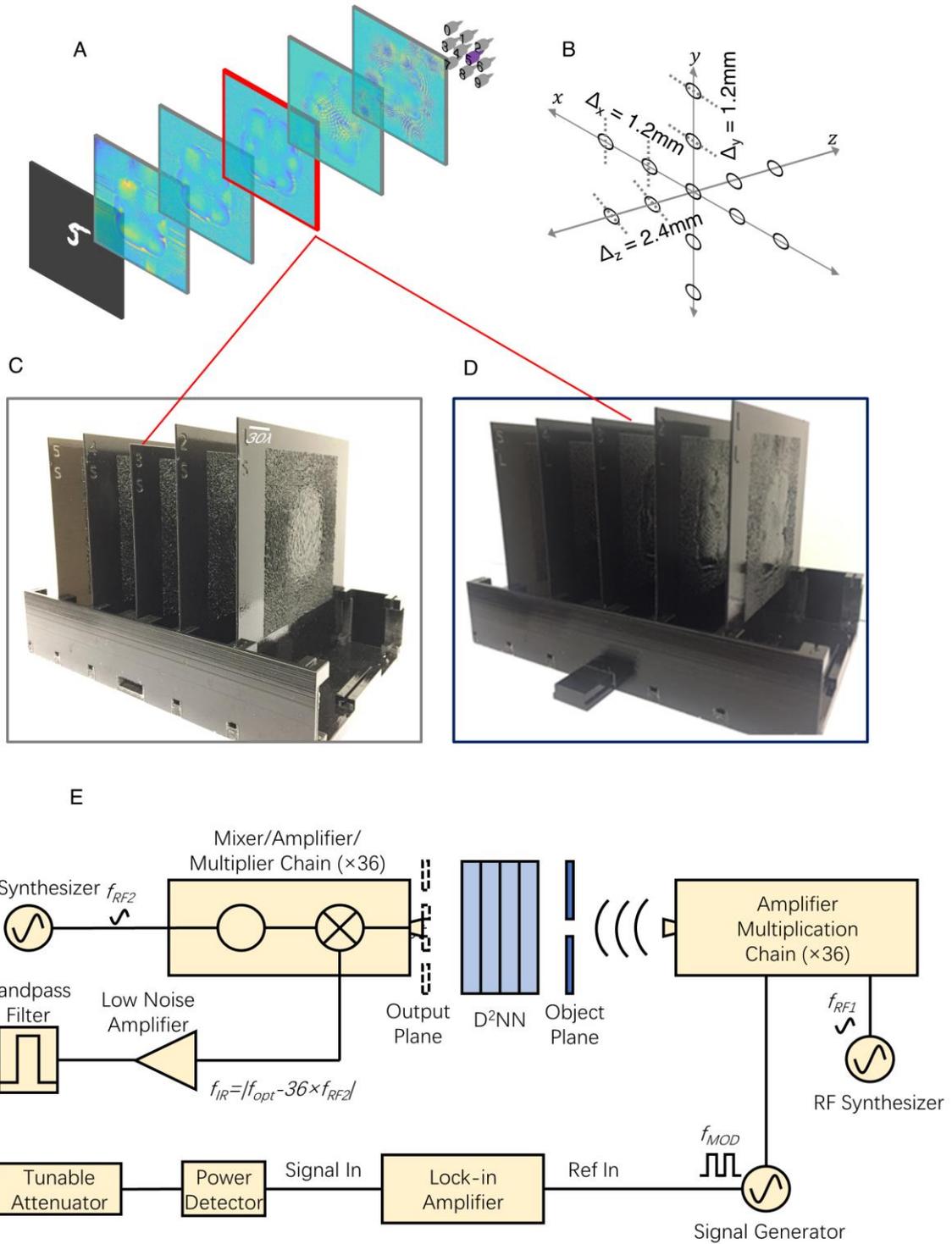

**Fig. 5: Experimental testing of v-D²NN framework. A** A diffractive optical network that is vaccinated against misalignments. This network is vaccinated against *both* lateral, $\Delta_{tr} = 4.24\lambda$, and axial, $\Delta_{z,tr} = 4.8\lambda$, misalignments. **B** The location of the 3rd diffractive layer was on purpose altered throughout our measurements. Except the central location, the remaining 12 spots induce an inter-layer misalignment. **C** The 3D printed error-free design shown in Fig. 1E. **D** The 3D printed vaccinated design shown in A and Fig. 1D. **E** The schematic of the experimental setup.

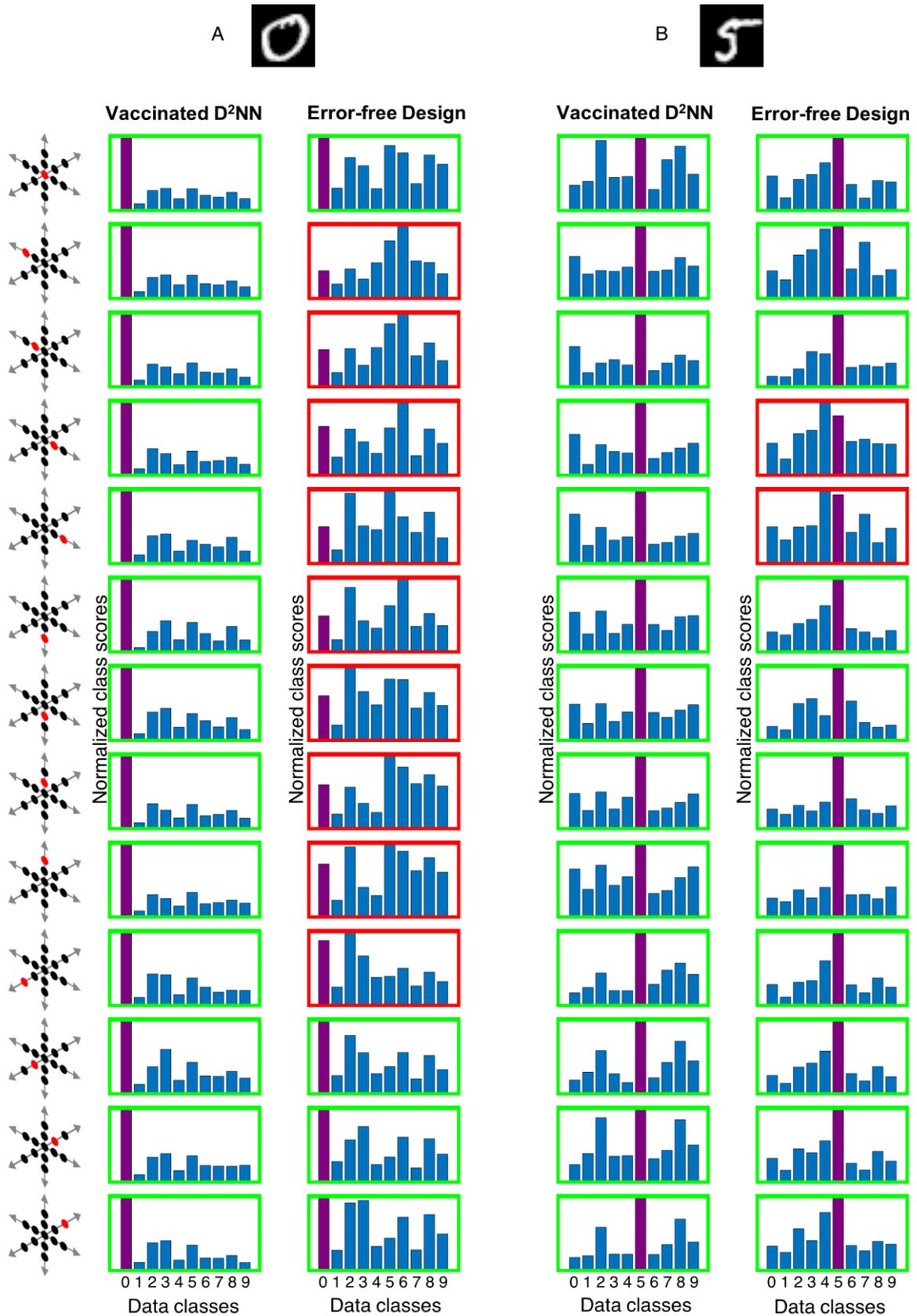

**Fig. 6: Experimental image classification results as a function of misalignments. A** The experimentally measured class scores for handwritten digit '0' selected from Set 1. **B** Same as A, except the input object is now a handwritten digit '5' selected from Set 2. The red dot within the coordinate system shown on the left-hand side represents the physical misalignment for each case (see Fig. 5B). Red (green) rectangles mean incorrect (correct) inference results. Refer to Supplementary Information (Figs. S2 and S3) for more examples of our experimental comparisons between these vaccinated and error-free diffractive designs.